\documentclass{elsart}
\def\be{\begin{equation}}
\def\ee{\end{equation}}
\def\bea{\begin{eqnarray}}
\def\eea{\end{eqnarray}}
\newcommand{\beas}{\begin{eqnarray*}}
\newcommand{\eeas}{\end{eqnarray*}}

\def\pdf{P_V}

\newcommand{\comment}[1]{}
\newcommand{\Avg}[1]{\left\langle{#1}\right\rangle}
\usepackage[english]{babel}
\usepackage{natbib}
\usepackage[dvips]{graphicx} 
\usepackage{color}
\usepackage{longtable}
\bibpunct[, ]{[}{]}{,}{a}{,}{,}

\usepackage{natbib}

\usepackage{amssymb}

\begin{document}

\begin{frontmatter}



\title{Decoding Information from noisy, redundant, and intentionally-distorted sources}


\author[yy,yz,yz,yz]{Yi-Kuo Yu,  Yi-Cheng Zhang, Paolo Laureti, Lionel Moret}

\address[yy]{National Center for Biotechnology Information,
 National Library of Medicine, \\
\hspace*{4pt}National Institutes of Health, Bethesda, MD 20894
}
\address[yz]{D\'epartement de Physique, P\'erolles, Universit\'e de Fribourg,
 CH-1700 Fribourg, Switzerland }

\begin{abstract}
Advances in information technology reduce barriers to 
 information propagation, but at the same time they also induce the  
 information overload problem. For the making of various decisions,
 mere digestion of the relevant information has   
 become a daunting task due to the massive amount of 
 information available. This information, such as that generated by 
  evaluation systems developed by various web sites, is in general
 useful but may be noisy and may also contain biased entries. 
 In this study, we establish a framework to systematically tackle
 the challenging problem of information decoding in the presence of
  massive and redundant data. When applied to a voting system, 
 our method simultaneously ranks the raters and 
 the ratees using only the evaluation data, consisting of an array of 
 scores each of which represents the rating of a ratee by a rater.  
 Not only is our appraoch effective in decoding information, it is also 
shown to be robust against various hypothetical types of noise as 
well as intentional abuses. 

\end{abstract}


\end{frontmatter}


\section{Introduction and Model}
With the rapid advances in {\it information technology}, and especially
the advent of the {\it internet}, information overload is becoming a
growing challenge for our society. In fact, daily life and
professional activities call for reliable information on a myriad of things,
and no individual is capable of knowing it all. We can, at best, rely
on other people's evaluations to indirectly form an assessment on a subject
 or item that happens to catch our interest. Numerous web sites have
already constructed evaluation systems which allow new users to
benefit from the feedback of previous users~\cite{ebay,E_commerce}.
However, even though many opinions can be found about any
single object (be it a book, a product, an idea),  
 they are frequently far from being consistent with each other, 
  perhaps because people are of different expertise and/or have different levels of 
 discernment. More often than not we are left without a clear, 
 definite answer. This situation calls for automated methods of collaborative
 information filtering and ranking~\cite{GroupLens}. 
 
Many web sites, which provide information filtering and evaluation for
 the general public, are themselves evaluated and ranked 
 by all the individuals via perhaps other web sites. 
 This {\it self-organized selection} has become increasingly 
 popular among internet users and may play an important 
  role in shaping the upcoming information-technology-mediated
  economics framework. Examples may be found through web sites
 such as {\it del.icio.us, www.digg.com, www.reddit.com, www.tailrank.com}
  et al. In a way, these sites all probe various selection and filtering 
 mechanisms with varying degrees of success. We believe that it is time  
 to examine the phenomena systematically in order to understand the
 theoretical foundation of information filtering.  

In this work we formulate a prototype model to cope with
such a challenge. Suppose $N$ users (raters)
 rate $M$ objects (ratees) in a given
category. Each user has an idiosyncratic rating capability 
 ($1/\sigma_i$ for rater $i$); each object has an 
intrinsic quality ($Q_l$ for object $l$). Both the rating capabilities 
and intrinsic qualities are assumed given and hidden. 
 We wish to find estimates, $q_l$ and $V_i$, as close as possible to 
 the hidden values, $Q_l$ and $\sigma_i^2$. Absent further information,
 people often use the simple arithematical average 
 $q_l \equiv \sum_{i=1}^N x_{il}/N$ as the estimate for $Q_l$, where
  $x_{il}$ is the rating assigned by user $i$ to object $l$.  
   With our additional assumption that users are of different 
 rating capabilities, we may regard the rating $x_{il}$   
 as the sum of $Q_l$  and a stochastic component 
 of typical size $\sigma_i$. Though many users report ratings 
 on a given object, these signals
  can be termed noisy since there is no sure way to tell which evaluation 
 is more reliable than the other. To make sense of these noisy signals 
 our only hope is to leverage the information redundancy and find the best 
 possible approximation to the hidden attributes.

As we will show below, the correctness of ranking can be distorted 
  when the true quality of each object is estimated by 
 the  na\"ive simple average of that object's ratings. This effect 
 is amplified especially when the typical $\sigma_i$s
 vary significantly so that the simple average may be biased 
by raters with large
 $\sigma$s. Our method, termed {\it iterative refinement}, can nevertheless 
 minimize the occurrence of this undesirable scenario. 


\section{Method}
Our task is to simultaneously 
 obtain good estimates $\{q_l\}$ and $\{ V_i \}$ 
 respectively for $\{ Q_l \}$ and $\{ \sigma_i^2 \}$ using 
 only the ratings $\{ x_{il} \}$.  
Absent knowledge of $\{ \sigma_i^2 \}$,  the simplest solution 
would be the  na\"ive arithematical average
$ q_{l} = {1 \over N} \sum_{i=1}^{N} x_{il} $. 
Though it conforms to the principle of `one person one vote',
such a na\"ive average is often said to suffer from the `tyranny of the
majority', especially when 
the majority is poorly informed. When one knows $\{ \sigma_i^2 \}$,
 one can improve the accuracy of the prediction 
  by giving more weight to experts with smaller $\sigma_i$s: 
\be\label{qalfa}
 q_{l}= \sum_{i=1}^{N}w_i x_{il}, 
\ee
where $w_i = w(\sigma_i)$ is a decreasing function of $\sigma_i$ and 
 with the normalization condition $\sum_i w_i = 1$. 
 In fact, if the $w_i$ are properly chosen so that Lindeberg's 
 condition~\cite{Feller} holds, $q_l-Q_l$ 
  becomes a zero-mean Gaussian random variable with variance
  $\propto 1/N$ when $N\to \infty$, thanks to the Lindeberg-Feller
 theorem~\cite{Feller,Chung}. Further, knowledge of $\{ \sigma_i^2 \}$ 
 actually allows the determination of optimal choice~\cite{Hoel} for the 
 weight $w_i \propto 1/\sigma_i^2$.

The problem, however, is that we know neither which are the best objects, 
 nor who are the best raters. Nevertheless, since a 
  better estimate of $\{ \sigma_i^2 \}$ will
 improve our estimate of $\{ Q_l \}$, we have devised 
 an iterative refinement method to simultaneously 
 extract the raters' rating capabilities and 
 the objects' intrinsic qualities. In particular, the rating capability of
 rater $i$ is estimated by
\be\label{sigma}
\sigma_i^2 \approx V_i\equiv \frac{1}{M}\sum_{l=1}^M (x_{il}- q_{l})^2 \;.  
\ee
 It is worth pointing out that due to error propagation 
 (e.g. estimating $Q_l$ by $q_l$), equation (\ref{sigma}) 
 is not the best possible estimator of $\sigma_i^2$. 
 Although it is possible to systematically compute {\it all} 
 the correction terms by creating effective Gaussian
 variables through recombining random variables,  
 we will not execute such a technique here to 
  avoid unnecessary complications. The more technical
  optimization of the method proposed will be discussed 
 in a forthcoming publication.

Assuming the weights to be of power-law type 
\be\label{weights}
w_i= \frac{1/V_i^{\beta }}{\sum_j 1/V_j^{\beta }}
\stackrel{q_l \to Q_l}{\Longrightarrow} \frac{1/\sigma_i^{2\beta }}
{\sum_j 1/\sigma_j^{2\beta }},
\ee
equations (\ref{qalfa}) and (\ref{sigma}) can be cast into a more
 suggestive form:
\bea
\sum_{i=1}^N {1\over {V_i^{\beta }}} (x_{il}-q_{l})&=&0, \;\;
l=1,2,\ldots,M;  \label{q_ave} \\
\sum_{l=1}^M {(x_{il}-q_{l})^2 \over
V_i}  &=& M, \;\; i=1,2,\ldots,N. \label{variance}  
\eea
 The above construction is intuitively
appealing especially when $\beta = 1/2$. 
Assuming that the ratings fluctuate around the hidden
qualities $Q_l$ with varying widths 
$\sigma_i$, 
the first equation is the sum of stochastic
variables with a unique, normalized width, and the second sum defines the
widths. On the other hand, the case $\beta = 1$ intimately mimics the
 optimal weighting choice~\cite{Hoel}.   

To solve for the $N+M$ unknowns using as many (nonlinear) equations is 
 usually achieved by casting the problem as a minimization~\cite{NR}. 
 This route is, in general, difficult for a nonlinear
  system because of the existence of multiple local minima that hinder 
 the finding of the global one. When the equations are such that
 local minima rarely occur, finding the solution becomes a  relatively
 straightforward numerical task. Fortunately, this is where our
 problem belongs. Our iterative refinement method
 starts with uniform weighting, then iterates
  eqs.~(\ref{q_ave},\ref{variance}) till convergence to the final solution
 with specified $\{ q_l \approx Q_l \}$ and $\{ V_i \approx \sigma_i^2\}$. 
Thus, we find simultaneously qualities of the ratees and raters.

As a cautionary note, we must comment that eq.~(\ref{q_ave}) with $\beta =1/2$ 
 takes a more gentle weight than suggested by the optimal 
weighting~\cite{Hoel}, which would recommend 
$\beta = 1$. We choose a softer weighting
 scheme to start because it enjoys a better 
numerical stability, as well as translational and scale invariance -- 
 the equations remain the same upon changing $q_l  \to c[q_l-g]$
 and $x_{il} \to  c[x_{il}-g]$.  The better numerical stability
 for $\beta = 1/2$ may be due to the fact that Lindeberg's 
 condition~\protect\cite{Feller} is always satisfied there. 
 Once the iterative procedure  
 starts, the weighting scheme is then shifted from $1/\sigma$ 
 towards $1/\sigma^2$ as the iterations progress. 
  If the Lindeberg's condition is satisfied for $1/2 < \beta \le 1$, 
 this construction guarantees the convergence of $q$ towards $Q$
  when each individual distribution function for $x_{il}$ is distinct
  but with a finite second moment,
 thanks to the Lindeberg-Feller Theorem~\cite{Feller,Chung}. 
 As will be shown, the correct convergence is obtained even when 
 the requirements leading to the general form of Law of
 Large Numbers are not satisfied, making our proposal far more 
 general than the traditional proven domain.

\section{Tests, results, and  analysis}
 Before any rating system can be put into real use, it must at 
 least pass theoretical quality control.  
The goal for a rating system is to produce the best 
approximation achievable, and to be robust 
against abuses and gaming attempts.
Any proposed decoding scheme should undergo testing under controlled
conditions ({\it i.e.} where $\{ Q_l \}$ is known), 
 whereas adaptation of a decoding method to realistic applications 
 usually requires a leap of faith since $\{ Q_l \}$ is unkonwn.
In fact, we would never know what the hidden, intrinsic attributes are
 or their underlying distribution. A decoding method has a higher chance
of success in the real world if it can consistently find the approximate hidden
attributes with a high precision under 
 a wide variety of controlled
conditions. We try to choose, among infinite possibilities, a number
 of case-studies we deem most significant.

We assume the intrinsic qualities ($Q_l$ for object $l$) 
 to be uniformly distributed and the ratings $x_{il}$
to be drawn from various individual distribution functions
 $\pdf(x_{il})$, centered around $Q_l$ and characterized by
 different widths $\sigma_i$. This implies that the ratings from 
 different users are assumed uncorrelated. Although we don't plan 
 to deal with this effect explicitly in this paper, we would like
 to point out that the correlation effect is automatically damped down 
 when using our iterative refinement. Because 
 we down weight users with weaker rating capabilities, sets of users with 
 poorer rating capabilities but having correlated ratings will have
  their votes downweighted and therefore won't be able to 
 bias the result much\footnote{This is assuming 
 that the ratings  of users are not all biased in the same
 direction for every object. If all the ratings are biased in the same
 direction, then a new consensus is formed and there is nothing 
 one can do to retrieve the correct attributes based only on the ratings
 given.}. For users with excellent
 rating capabilities but having correlated ratings, keeping or removing 
 the redundancy does not produce much effect on the final
 result either. The correlation between users, therefore, does not
 have a prominent effect.  
   We will thus present the study of the correlation effect in 
 a separate publication, and in the mean time return to
 the case of uncorrelated users.   

To be specific, we shall employ the following voting distributions:
\bea
\pdf(x_{il}) &= &{1\over \sqrt{2}\sigma_i} 
e^{-\sqrt{2}|x_{il}-Q_l|/\sigma_i} 
\label{V.Laplace} \\
\pdf(x_{il}) &= & {(f-1)/2 \over \sqrt{ C_f \sigma_i^2/2 }}
 \left( {|x_{il}-Q_l| \over \sqrt{C_f \sigma_i^2/2 }}
 + 1 \right)^{-f} 
\label{V.Powerlaw}   
\eea
 where $C_f = (f-2)(f-3)$ for $f > 3$. In this case
both distributions have finite second moments given by 
$\int \pdf(x_{il}) [x_{il}-Q_l]^2 dx_{il} 
 =   \sigma_i^2$. We may extend the exponent $f$ to be in the range
$ 1<f\le 3 $ at the expense of a divergent second moment,
and set $C_{f\le 3} = 2$. The resulting distribution (\ref{V.Powerlaw}),
falling outside the realm of the central limit theorem, will also be considered.
Finally, the distribution widths  $\sigma_i$ are also randomly 
drawn from a distribution function $p$. 
The broader the distribution function, {\it i.e.} the greater the inhomogeneity
in rating capabilities, the harder it is to have resulting $q_l$'s
close to the intrinsic $Q_l$'s. 

 As a quantitative measure of the accuracy of 
 any estimation method, we use a Euclidean-like distance between
the estimated solution $\{q_l\}$ and the intrinsic values $\{ Q_l \}$: 
\be \label{measure_dist}
\Delta=\sqrt{{1\over M}\sum_{l=1}^M (q_l-Q_l)^2}.  
\ee
Since in many applications it is required to rank the $M$ objects in
  order of decreasing quality, it is useful to compare the estimated
  ranked list $L_E$ with the intrinsic one $L_I$.
 A measure $I(p)$ is introduced to examine the ranking integrity  
within the top $p$ proportion of the
 object's quality list:
\be \label{measure_rank}
I(p) \equiv {1\over [pM]} \sum_{R=1}^{[pM]} {n(R) \over R},
\ee   
where $[x]$ indicates the maximum integer that is smaller than 
 or equal to $x$.  
Here $n(R)$ denotes the number of objects, among the top $R$ ones in
 the estimated ranking list $L_E$, whose 
 intrinsic qualities rank among the top $R$ in the intrinsic list
 $L_I$. 
The higher the quantity $I(p)$, the 
 better the overlap between the estimated ranking and the intrinsic one. 
 For illustrative purposes, we shall consider $p=0.5$ in this
 study. It is worth pointing out that the expectation value
 of $I(p)$ from random sampling can be calculated (see supplementary
 information for the complete proof):
\be \label{I_bg}
\langle I(p) \rangle_0 = {[pM]+1 \over 2M}\;\;
 \stackrel{M \gg 1}{\longrightarrow} \;  p/2;
\ee
In order to have a more precise measure of the variation of 
 $\Delta$ and $I(p)$ around their respective means, 
 we calculate the ``up" variance
 $\langle (\Delta - \langle \Delta \rangle)^2\rangle_{\Delta >\langle \Delta 
\rangle }$
 and ``down" variance $\langle (\Delta - \langle \Delta 
\rangle)^2\rangle_{\Delta < \langle \Delta \rangle }$ 
(identically in $I(p)$ case) and report their square roots 
 as asymmetric error bars 
in Figs. 1, 3 and 5.

\subsection*{Effect of the number of objects and raters}
Both the distance measure 
$\Delta$ in eq.~(\ref{measure_dist}) and the ranking-integrity measure
 $I(p)$ in eq.~(\ref{measure_rank}) are  
 functions of $M$ and implicitly of $N$. 
 For simplicity we will assume $M=N$ and measure both  $\Delta$ 
 and $I(p=0.5)$ for various $N$ and different stochastic distribution 
 functions $\pdf(x)$. 

\begin{figure}[ht]
\begin{center}
\includegraphics[width=0.58\columnwidth,angle=-90]{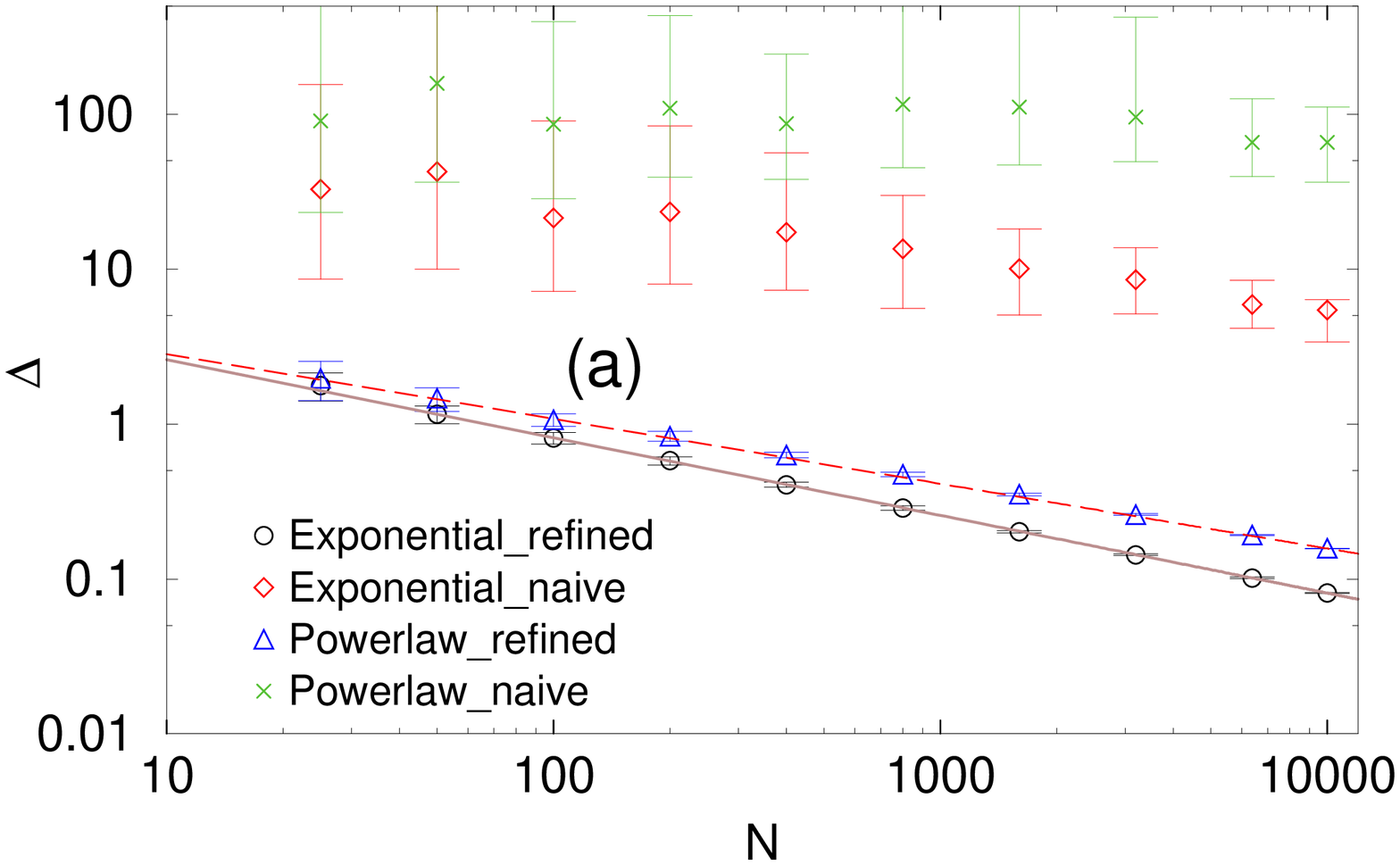}\\
\includegraphics[width=0.6\columnwidth,angle=-90]{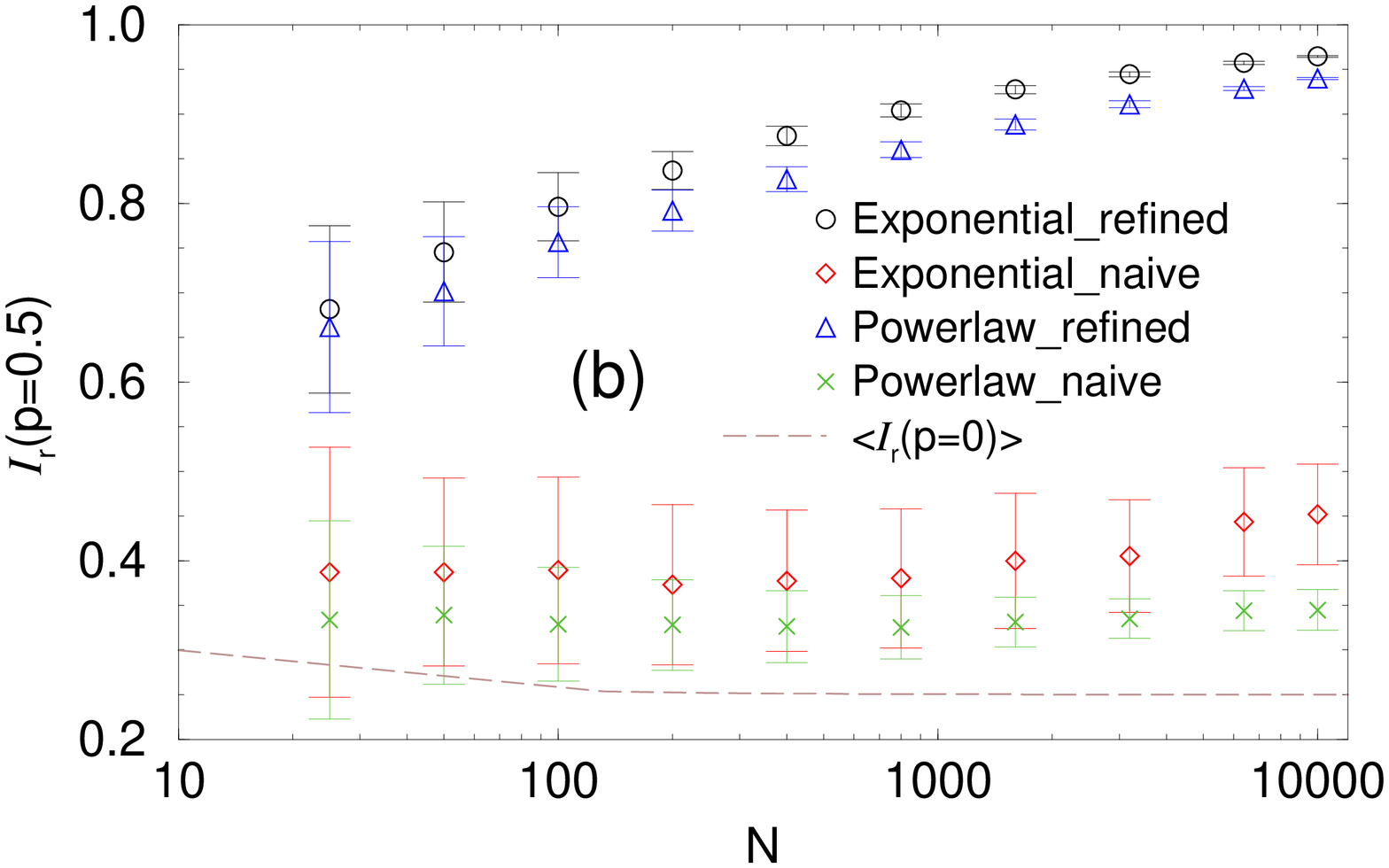}
\vspace{-0.1in}
\end{center}
\caption{Figures (a) and (b) show, respectively, the dependence 
of $\Delta$, from
eq.~(\ref{measure_dist}), and of $I(p=0.5)$, from
 eq.~(\ref{measure_rank}), on the system size $N$. 
 The intrinsic quality of each object is 
 randomly uniformly drawn from the interval $(0, 10)$. 
A small finite constant $\epsilon = 0.001$ is introduced so that 
 the judging width
is always bounded by ${\rm const}/\epsilon$
  to avoid the potential artifact of infinite width.  
 For the power-law case (eq.~(\ref{V.Powerlaw})), we picked $f=2.5$,
 so that the second moment
 of $V$ diverges, to test the robustness of our method. Indeed, 
the deviations $\Delta$ still decrease steadily 
 with size, although not as fast as for the exponential voting
 distribution (eq. (\ref{V.Laplace})), while the ranking integrity grows.
 The solid (dashed) line passing through the exponential (power-law)
  voting distribution in graph (a) has slope $-0.50 \pm 0.01$ 
 ($-0.42 \pm 0.01$). The dashed line in graph (b) 
 is the result for the random case (eq. (\ref{I_bg})). 
 Error bars have been estimated from 
 the standard deviation of
one hundred realizations 
 of different initial configurations. 
}\label{fig1}
\end{figure}

As shown in Fig.~\ref{fig1}(a), using the iterative refinement method   
 the difference between the $q$'s and the $Q$'s becomes smaller steadily 
with increasing $N$. When using the exponential 
 distribution (\ref{V.Laplace})
 for voting, within numerical errors 
 the downward slope equals $-1/2$ as expected from the Law of  Large Numbers. 
 When the voting distribution function has a power law tail  
that prevents a finite second moment, the proposed method still shows 
  steady improvement with increasing $N$.  In Fig.~\ref{fig1}(a) we
 also show, as a comparison, the
simple average with equal weights: the difference $\Delta$ is much larger and
its convergence to zero is not guaranteed. 
Moreover, the precision for a moderately large set of data is strengthened
by the  rapid decrease of the error bars.    

In Fig.~\ref{fig1}(b) we show how the ranking integrity changes with size. 
 There is a clear separation between the na\"ive arithematical average and the
 iterative refinement method. 
 The increase of the ranking integrity with $N$  
 confirms that the robustness of our method increases steadily
 with the system size.
 When using the naive averages, on the other hand, it seems to change
 only slightly or not at all, within the precision of the indicated
 error bars.  

\subsection*{Self-evaluating community} 
Our  method can be easily
applied to another context: a self-evaluating community. Suppose 
a community of experts tries to find the intrinsic
ranking of its own membership. 
Each expert has an opinion on every other member in this
community and opinions are uncorrelated. 
In this case we have $N=M$ and the asymmetrical matrix element $\{x_{ij} \}$
denotes $i$th member's rating on $j$th member. Thus, each member has a
given attribute we call quality as well as a given rating capability. 
With minor modifications, (\ref{q_ave}) and (\ref{variance}) become 
\bea
\sum_{i\ne j} {1\over {V_i^{\beta }}} (x_{ij}-q_{j})&=&0, \;\;
 j =1,2,\ldots,N \; ,   \label{q_ave_self} \\
\sum_{j\ne i} {(x_{ij}-q_{j} )^2\over
V_i} &=& N, \;\; i=1,2,\ldots,N \; . \label{variance_self}  
\eea 
Our iterative approach can be readily used to find the
solutions for the $q$'s and the $V$'s (estimated for
 $Q$'s and $\sigma^2$'s). There are members who are judged by
others as higher quality authorities and some members turn out to excel
at rating fellow members. Still others are good at both.

\begin{figure}[ht]
\begin{center}
\includegraphics[width=0.72\columnwidth,angle=-90]{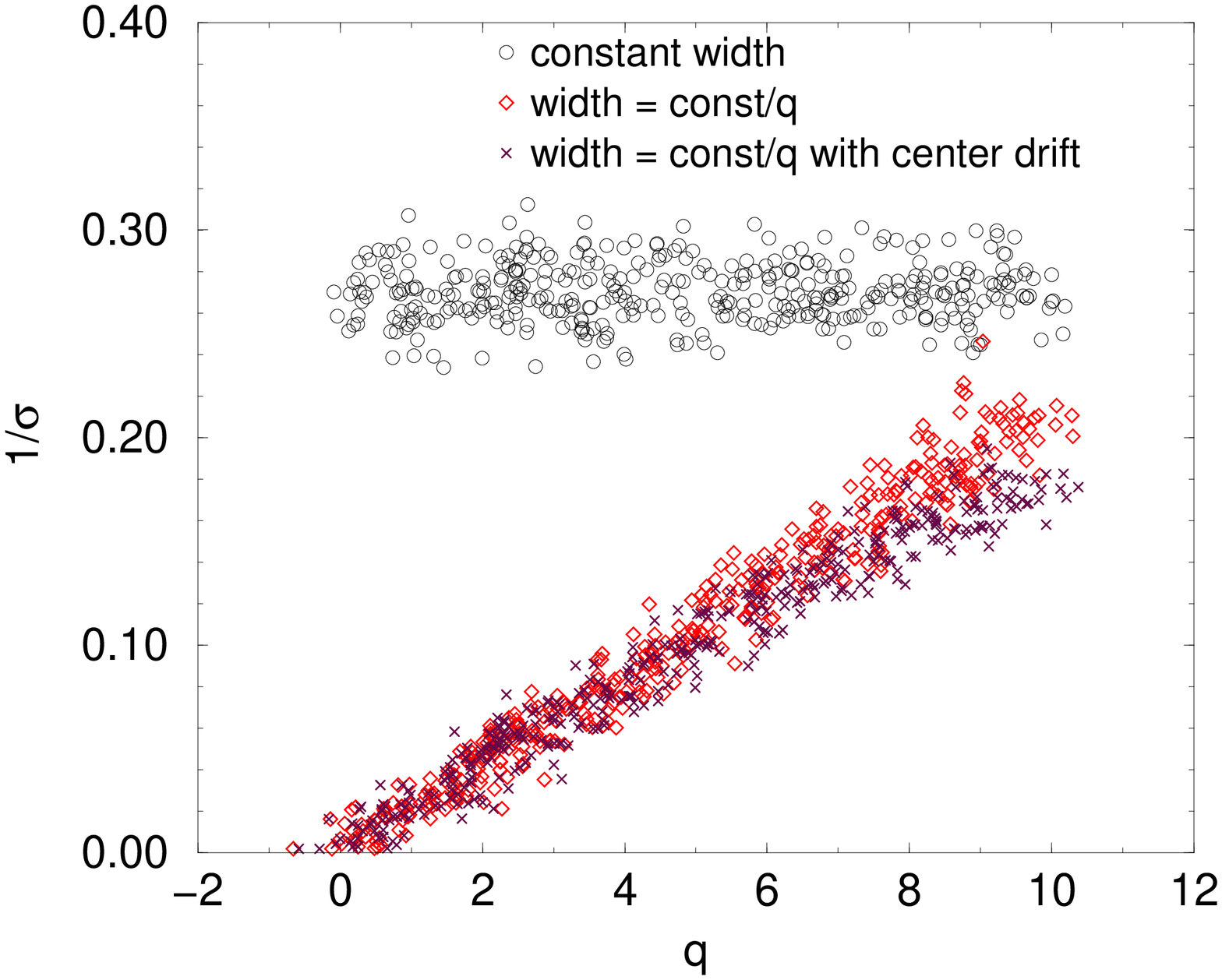}
\vspace{-0.1in}
\end{center}
\caption{Experts' estimated rating capability versus estimated quality. 
Using $N=400$, data is taken
 from the outcome of a typical single run with, respectively,
 $\sigma_i={\rm const.}$ (circles), $\sigma_i={\rm const.}/Q_i$ with a
 random center-drift uniformly distributed between $[-5,5]$ ({\sf x} signs) 
 and without drift (diamonds).  
}\label{fig2}
\end{figure}

For the self-evaluating community,  
the simulation results on $\Delta$ and $I(p=0.5)$ largely 
 agree with those presented in Fig.~\ref{fig1}   
 Apparently, one may wishfully believe that there exist correlations between 
 the qualities of being good experts and being good raters. 
 For example, in the real world people often assume that being 
 a good expert automatically implies being a good rater.
 However, we should be cautious in this regard:
though very likely the two types of quality are somewhat correlated,
we should let the data itself bring out evidence which may 
 support or undermine such a hypothesis. To demonstrate this possibility, 
 we have run a simulation where everyone has the same rating capability 
 and another where the rating capability $1/\sigma_i$ of user $i$ is 
 directly proportional 
 to his quality $Q_i$ of being a good expert. Although no information about the
  correlations between $Q_i$ and $\sigma_i$ is fed into 
 our iterative refinement method, the result shown in 
 Fig.~\ref{fig2} -- plotting  $1/V^{1/2}$ versus $q$ 
--  reflects strongly
  the respective underlying correlations. 

\subsection*{Individualized biases of the target $Q_l$} 
One may object to the fact that each rater's distribution is
 symmetrical around the intrinsic attribute.
To subject our method to more severe tests, 
we now relax this condition: we allow each rater
to have an individual distribution function not only with a specific
width, but also an object-dependent individualized biased center. Thus,  
 $x_{il}$ is drawn around $Q_l+\Delta Q_{il}$, 
 where the center drift $\Delta Q_{il}$ represents the individual bias. 

For testing purposes, the quantity $\Delta Q_{il}$ was drawn 
 from a uniform distribution inside $[-5,5]$, while the intrinsic qualities
 $Q_l$ are within the range $[0,10]$. Despite the fact that
 $\Delta Q$ has a rather large range, the convergence of $q$ to $Q$ 
 is almost as good as in the unbiased case. There 
 is only a small increase in $\Delta$ and a negligible decrease
  in $I(p=0.5)$. Even in the case of the self-evaluating system,
 the modification does not spoil the underlying characteristics. As shown in
 Fig.~\ref{fig2}, when the rating capability is directly proportional to
 the expert quality, the linear relationship found between 
 $1/V^{1/2}$ and $q$ 
 still holds, except with a smaller slope.

We may conclude that the proposed method is quite effective in decoding the
hidden attributes in the controlled numerical experiments. However,
before proposing it for real applications we must face another type of
challenge: members may harbor private agendas and may willfully distort
information. So far we have dealt with random, neutral noise,
which we shall call the first kind. The second kind of noise, is unique
for intended human actions. Comparing with the information theory
 of Shannon for transmitting signals via a noisy channel~\cite{shannon},
 which by definition
deals with noise of the first kind, we must be wary that our method
should be relatively robust against willful distortion as well. As people often
observe in real life information collection and evaluation,
gaming the system is often hard to detect, and still harder to stop.
We now turn to this case.

\subsection*{Intentional distortion}
Consider the context of a mutually evaluating community. Since 
friendships and rivalries are as ancient as civilization in any human grouping,  
 we must expect that some will give a more favorable evaluation for their friends,
with upward deviations often larger than what their rating capability would
warrant. Likewise they may rate down their enemies. For this reason we
shall extend the proposed model to include some friendship-enemy pairs
 among the members of the community. A simple way to implement this is
 to pick in
turn every member, sending out a fixed number of friendly links landing
randomly on fellow members, and the same number of enemy links. 
 As a result of this procedure, 
any member can receive more or less friendly or hostile links. When two
members are linked by a friendly link, they trade favors by up-rating
each other by a upwards bias. Likewise, two enemies will rate down each
other by a downward bias. To simulate this effect, when a member $i$ votes
 on his friend (enemy) $j$, we increase (decrease) the vote $x_{ij}$
 by $2\sigma_i$.  As a control parameter we  denote by $\gamma$ the
percentage of friendly and hostile links. For instance, at $\gamma=30\%$
each member has $30\%$ out of the total fellow members as friends,
and as many enemies. Thus, the remaining $40\%$ are neutral ones to him. 

\begin{figure}[ht]
\begin{center}
\includegraphics[width=0.58\columnwidth,angle=-90]{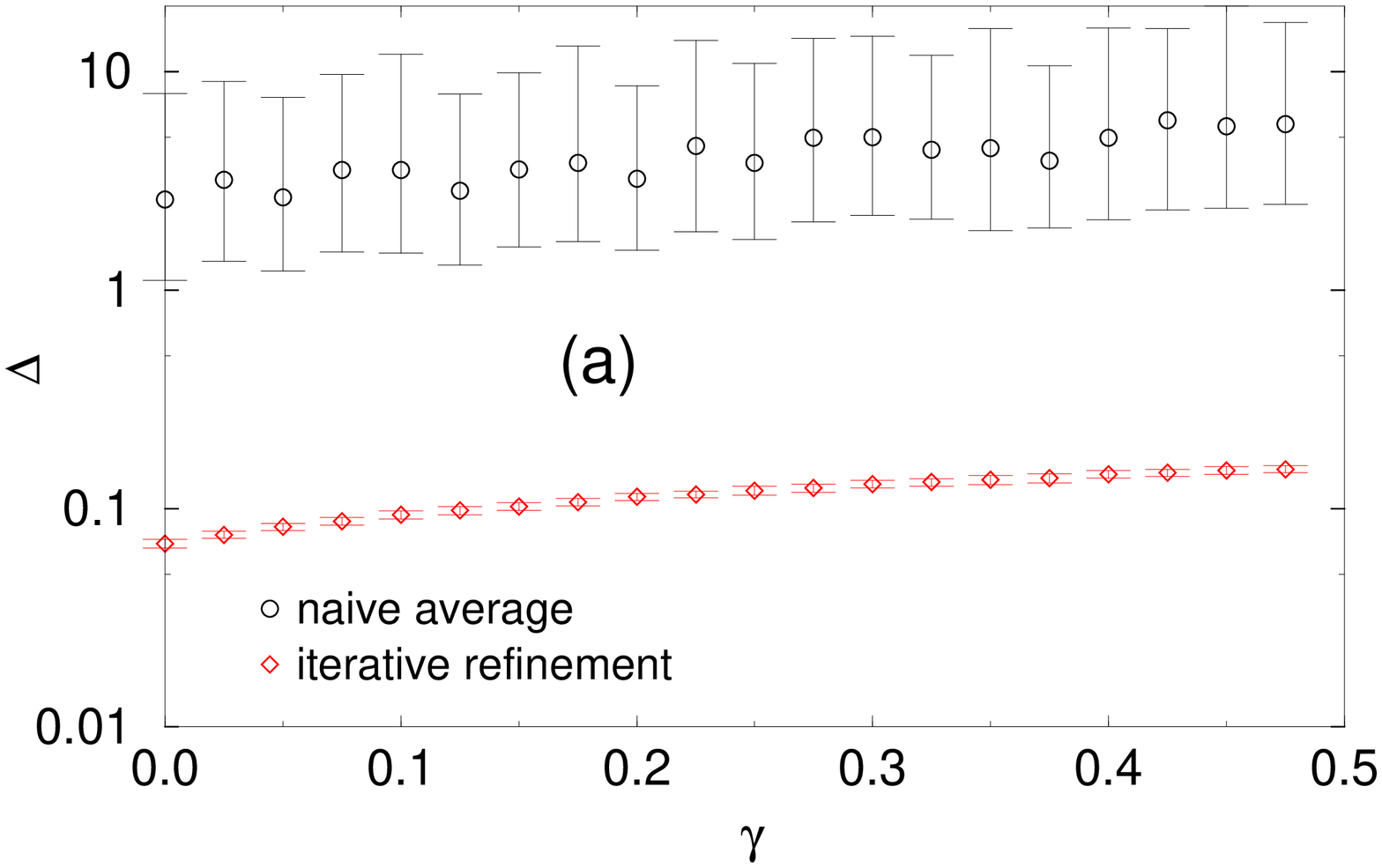}\\
\includegraphics[width=0.6\columnwidth,angle=-90]{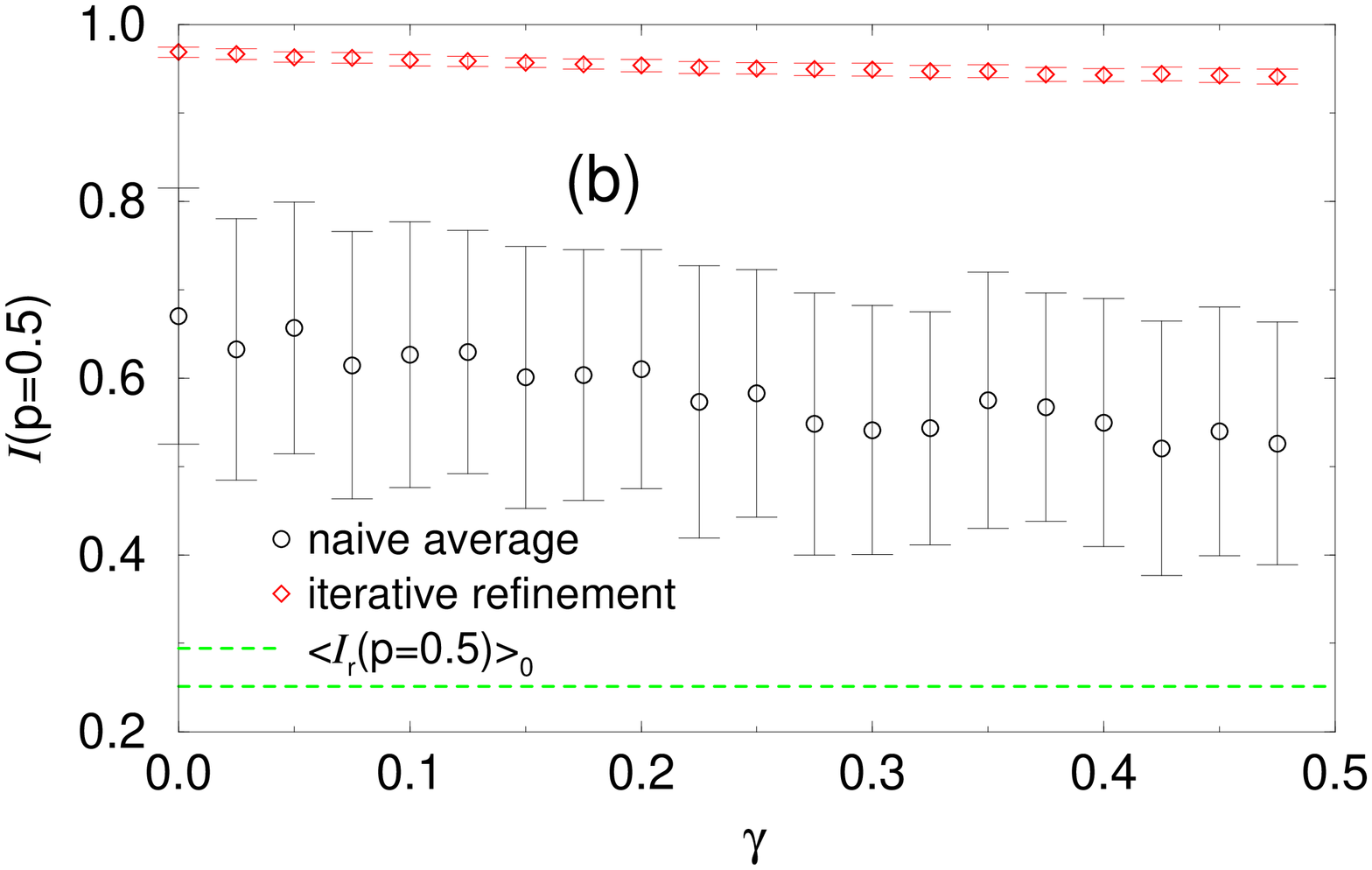}
\vspace{-0.1in}
\end{center}
\caption{The effect of intentional distortion on $\Delta$. 
The abscissa records $\gamma$, the ratio 
 of friends and enemies to the population in a community;  
  the ordinate documents the distance $\Delta$ (a) and the rank
 integrity $I(p=0.5)$ (b).  
 Using $N=400$, the data are obtained 
 from the outcome of one hundred simulation runs. 
 The standard deviations from the mean value are shown as a vertical
 bar around the average value. 
}\label{fig3}
\end{figure}

In Fig.~\ref{fig3} we see that, as the percentage $\gamma$ increases, i.e. 
the community becomes more and more corrupted and 
 ratings become less and less fair, the decoding
efficiency deteriorates. Specifically, one should note the increase
 of $\Delta$ and the decrease of $I(p=0.5)$ as $\gamma$ increases. 
However, the overall efficiency holds remarkably
well in the face of the massive information corruption. Even when the
majority of fellow members are either friends or enemies, the
solution $q$ still remains very close to $Q$. As a comparison, we see that the
 ranking integrity from the 
simple average quickly worsens and comes close to the expectation value
 from informationless random sampling.

When a member rates another member far away from the intrinsic attribute,
the mischief costs him somewhat in credibility $1/V^{1/2}$, 
 which is the estimate of his rating capability $1/\sigma$. 
 If there are
a high fraction of friends and enemies, then all are adversely affected
in their rating capabilities; this is best seen from Fig.~\ref{fig3}(b),
 which
 shows how the ranking integrity worsens as the friend/enemy fraction
  $\gamma$ increases. 

\subsection*{Stability against worst abuses}
It is instructive to examine the maximum ability 
for a member to willfully
distort information about another. 
 We would like to investigate the effect of a willful distortion 
  on the final rating of the targeted member as well as
 on the cheater's estimated rating capability, which can
 also be interpreted as his credibility within the community. 
 For this purpose, the simplest method is to consider a fair community, 
 where only one member harbors a hidden agenda to distort the rating
  of another member by a very wide margin. 
 Since all other raters are fair, the impact of this distortion
can be calculated, as well as the repercussion on his own rating capability,
judged by the community. If the cost is high for the cheater compared
to the possible impact the cheating would have, then we may conclude
that the method is naturally robust in constraining cheating behavior; if
the cost is low for a similar impact, we should expect that cheating would
run rampant and additional features are called for to prevent it.

It is practical to represent all the members in two  rank lists: one for
their quality judged by the fellow members; another for their rating
capability as the result of their behavior in judging others. This is
similar in spirit to the notion of authority and hub \cite{Kleinberg}, 
two qualities that are also central in ranking web pages. 
 Assume one cheater in an otherwise fair community. 
 Because promotion and demotion have almost identical costs on 
 the cheater's rating capability, we need illustrate in detail 
 only one of the two possible ways of cheating. 
 Suppose the cheater promotes a friend beyond the 
 intrinsic merit by a quantity $B$. 
We wish to know ($i$) by how much this promotion 
 would move  his friend up in the attribute rank list, and ($ii$) 
 by how much this cheating would move the cheater down on the rating 
 capability rank list. 
We may further inquire what is the maximum distortion a member
can possibly create. This is interesting since only by knowing the worst
case scenario can we learn how robust this method is. 

First we note that there indeed exists 
an upper bound on the possible cheating. A member launching
a desperate distortion act, not caring about any damage to his own
rating capability, could not favor his fellow indefinitely. 
Because the rating from cheater $I$ is weighted by 
$1/V_I^{{\beta } > 1/2}$ (see the METHOD section) 
 and $V_I(B\gg 1)$ is related to 
$V_I(B=0)$ via the relation\footnote{This relation stands valid
 even when $B \approx 1$ except that then there are other 
 correction terms of comparable order. A detailed study of this effect
 will be presented in a forthcoming publication.  }
$$
V_I(B\gg1)-V_I(B=0) = \frac{B^2}{N}[1+ \mathcal{O}(1/B)], 
$$
 the overall contribution from cheater $I$ on his pal $J$ that he is 
 trying to promote behaves like  
$$
 \approx  {x_{IJ}+B \over [\sigma_I^2(B=0) + \frac{B^2}{N} ]^{\beta}}
$$
 for large $B$, and it becomes vanishingly small when $\beta > 1/2$. 
 This indicates that when rating an object way out of proportion,
 the net effect is as if the distorted vote never existed and
 our desperate cheater may not want to inflict such
an egregious distortion lest his credibility
drops to the bottom of the list. Such a desperate act 
 in reality inflicts the maximal damage to his own
credibility while achieving little desired result. 
Therefore, a rational cheater may then take a more calculated approach to  
 produce a maximal distortion.  A highly credible member can generate a larger
 distortion than an average member, should he choose to do so;   
a member on the bottom of the rating capability rank list has, 
 however, weaker impact on promoting or demoting other members. 

 From the system point of view, we do not have to
suffer from the maximal distortion. It is easy
to detect such attempts and the cheater's rating can be simply ignored
and, at our discretion, the cheater's rating capability can be restored
since his contribution evaluating other neutral members is a valuable
service we want to keep. We call this the ``{\sf S}'' strategy. 
The implementation of the ``{\sf S}'' strategy is flexible, we
can either choose to inflict a punitive penalty or simply detect cheating
and ignore it. Although it should depend on how reliably we can 
 detect cheating, we currently implement the ``{\sf S}'' strategy in 
 two steps: if the rating from voter $i$ to object $l$ is more than 
 two $\sqrt{V_i}$ away from $q_l$, we down weight the rating $x_{il}$ by an
 additional factor  $(2\sqrt{V_i}/|x_{il}-q_l|)^{1/2}$ 
 and we totally discard the weight 
 whenever $|x_{il}-q_l| > 3\sqrt{V_i}$. 

\begin{figure}[ht]
\begin{center}
\includegraphics[width=0.6\columnwidth,angle=-90]{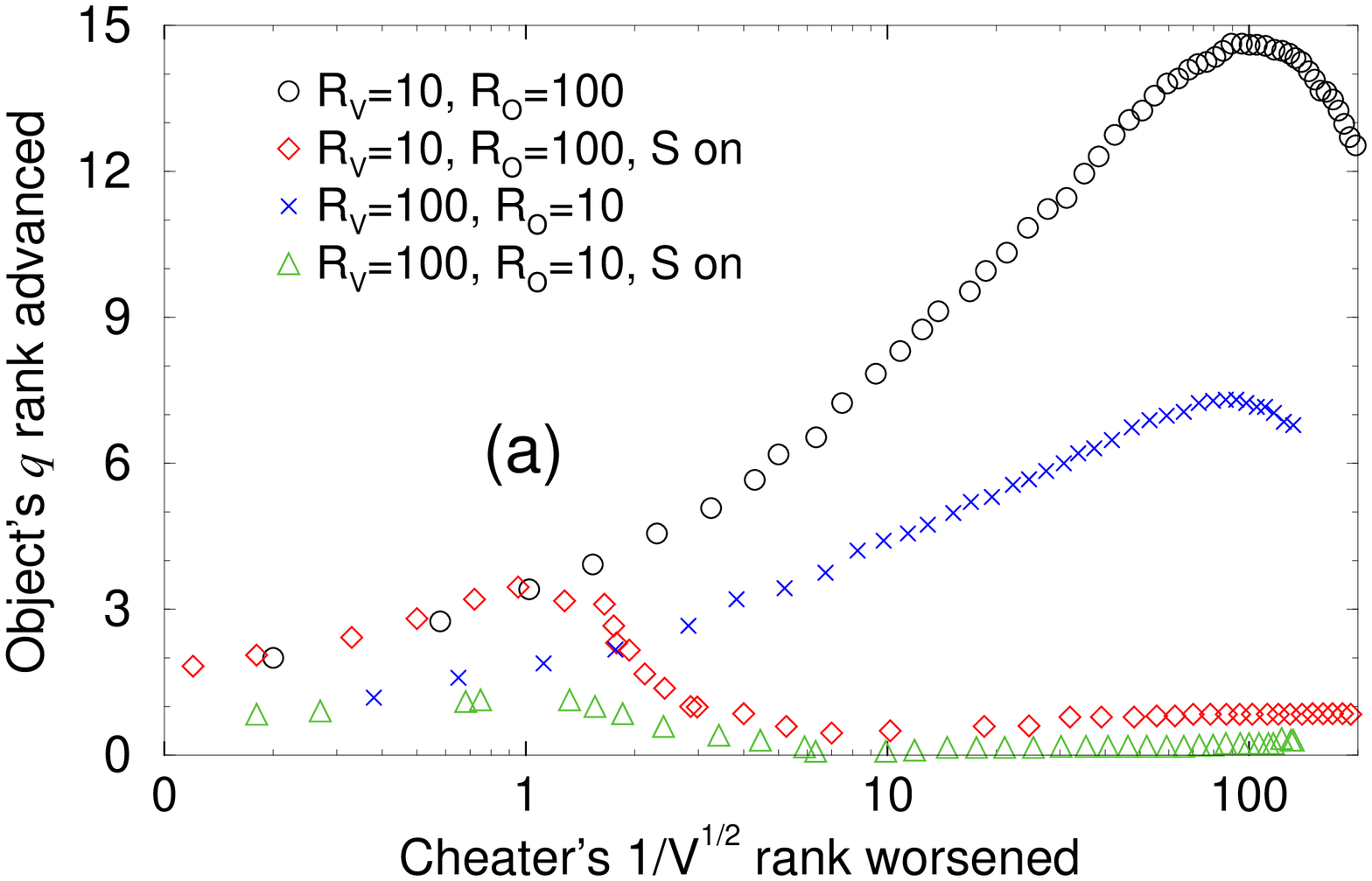}\\
\includegraphics[width=0.6\columnwidth,angle=-90]{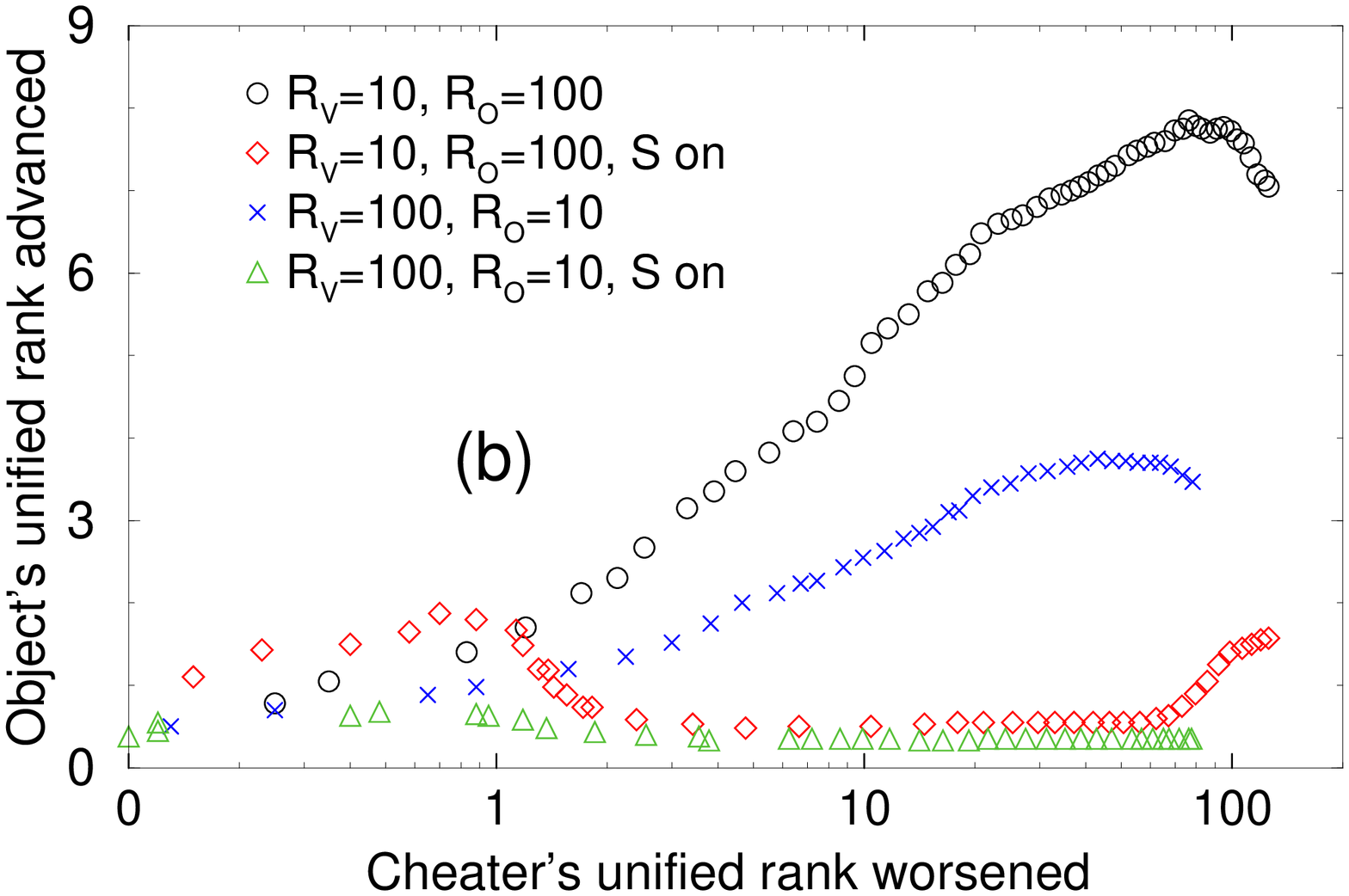}
\vspace{-0.1in}
\end{center}
\caption{The effect of extreme intentional distortion on the object's
 rank (a) and on the object's unified rank(b). ${\sf R_V}$ represents the 
 cheater's rank in the intrinsic rating capability list, while ${\sf R_O}$ 
represents the benefited object's rank in the intrinsic quality list. 
  By increasing the
 amount of intentional distortion $B$, the $1/\sqrt{V}$ rank of the cheater 
 worsens while the $q$ rank of the benefited object also covaries.
 When the ``{\sf S}'' strategy is turned on, we see that the cheater
 has very little impact in distortion, despite how willing he is 
 to sacrifice himself.   See the text for the explanation of the 
 final bump in the red curve of figure (b).  
}\label{fig4}
\end{figure}

 A simulation of $400$ experts in 
 a self-evaluating community is performed to test the effect of
 intentional distortions. In this simulation,  we assume that each 
 rater's  rating capability is directly proportional to his intrinsic
 quality. Forty realizations of different rating matrices are 
 evaluated with the iterative refinement approach and the rank changes 
 due to cheating for each rating matrix are averaged. 
 We note that, as summarized in Fig.~\ref{fig4}(a),  
 when the cheater's intrinsic rating capability ranks high, he 
 can promote others more than a cheater with 
 mediocre rating capability. 
 The cheater will have to pay a price of moving his rank down on
 the rating capability list by about $100$ to achieve the maximum 
 distortion of about $10 - 15$. However, once we turn on the 
 ``{\sf S}'' strategy, appreciable distortion can no longer 
 be achieved, as demonstrated in the bottom two curves of 
Fig.~\ref{fig4}(a).     
 Therefore, it is possible to maintain a high degree of fairness
 and discourage cheating when this new method is employed in real 
 society. With more information shared and less cheating allowed, 
 our society can grow into a happier and healthier whole.    

In real life the rating capability and one's intrinsic
 quality might not be always correlated, as plausibly 
 assumed in our simulation.  However, without
any presumption, working with real data may well reveal any correlation,
since the method we propose does not exclude any specific one. We may propose
a combined quality parameter to represent a member's overall capability,
$q/\sqrt{V}$. Any member, found to rank high on this new combined rank list,
will be both judged highly by fellow members and behave well
in judging others. The new rank list would also serve as a deterrent to
cheating: any willful distortion attempt would cost a cheater somewhat
in overall quality. The cost-benefit analysis now becomes even simpler:
cheating may move up (down) a friend's (enemy's) rating, but the cheater's
own overall rating slips. In the previously mentioned 
 simulation, we also document the change of this unified 
 rank for both the cheater and the benefited  object.  
 In Fig.~\ref{fig4}(b) it appears again that cheating does not pay. 
 Note that the bump in the tail of ${\sf R_V=10,~R_O=100}$, with 
strategy ``{\sf S}''
 turned on, is 
 not an indication of the malfunction of our method. In fact, there 
 the advance of the object's unified rank is due to the 
 fact that the cheater's unified  rank has dropped below that of the object's.
  This nice feature again discourages severe cheating!

\begin{figure}[ht]
\begin{center}
\includegraphics[width=0.95\columnwidth,angle=0]{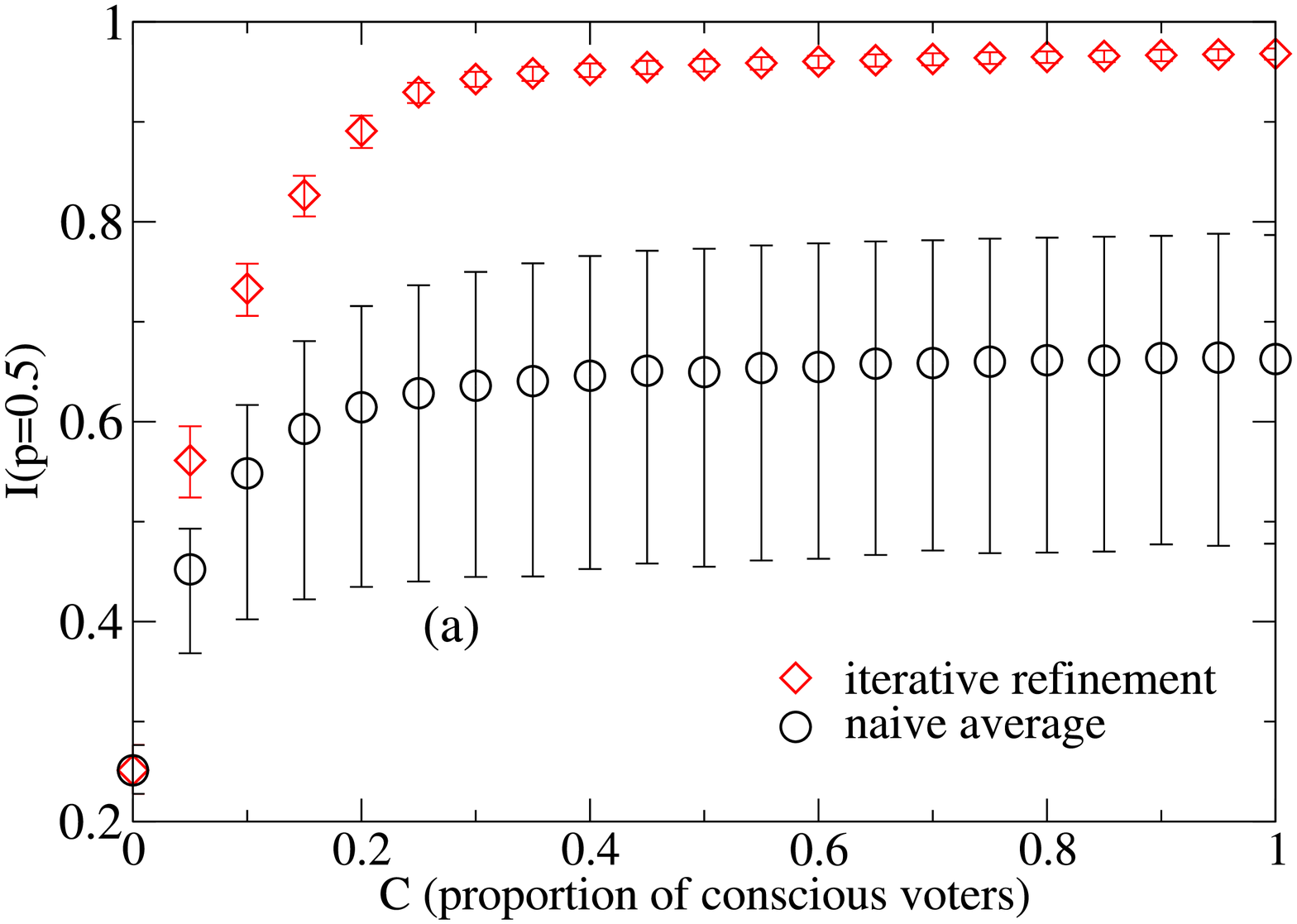}\\ \vspace{-0.12in}
\includegraphics[width=0.95\columnwidth,angle=0]{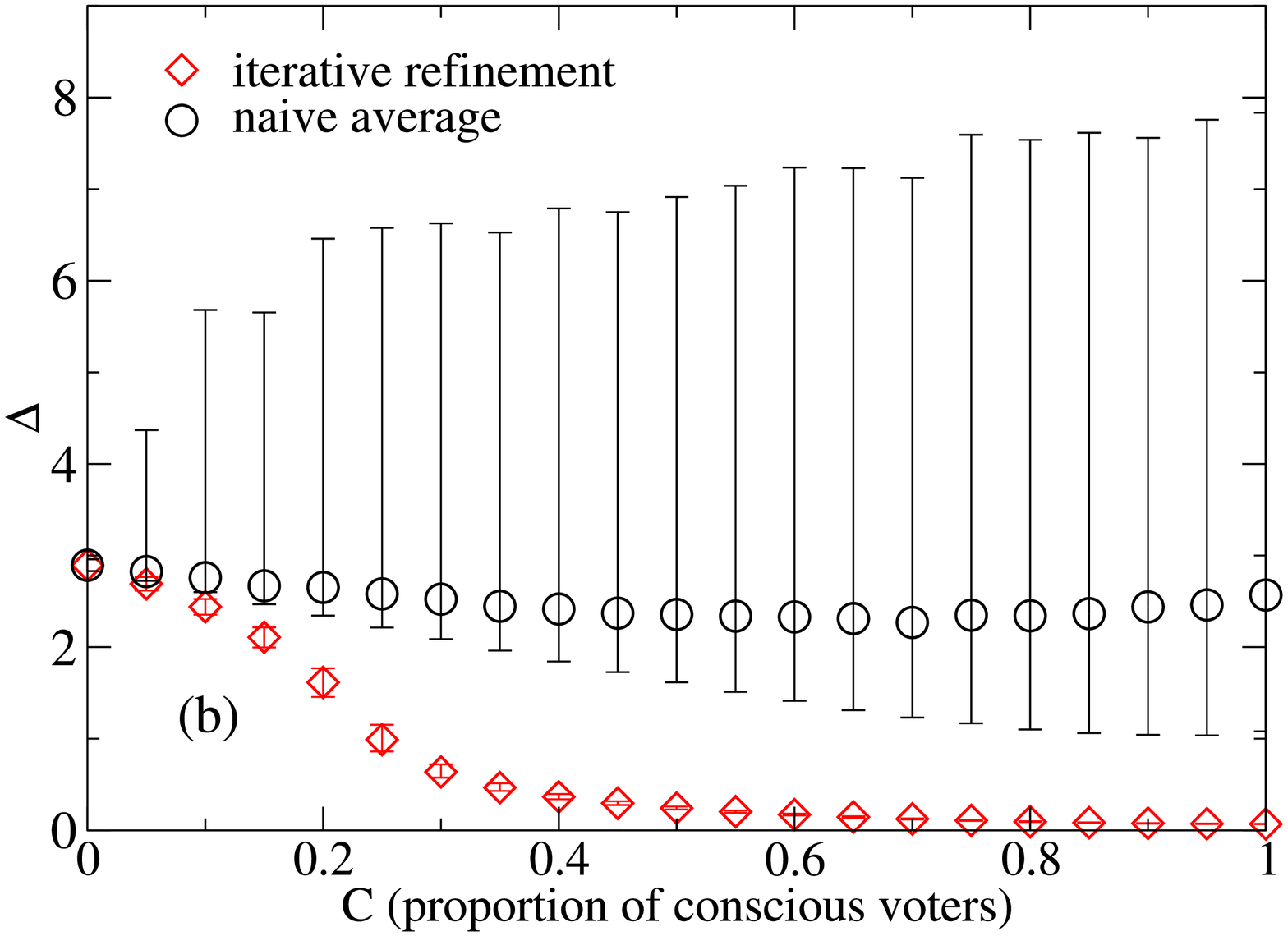}
\vspace{-0.1in}
\end{center}
\caption{The effect of ignorance.  $C$ represents the 
 proportion of conscious raters who vote according to the exponential
 distribution (\ref{V.Laplace}).  
(a) Ranking integrity $I(p=0.5)$ against $C$. (b) $\Delta$ against $C$.
We see that $I(p=0.5)$ (resp. $\Delta$)  increases (decreases) rapidly,
 and becomes much bigger (smaller) than the simple average, 
 when $C > 0.4$. $\sigma$'s of conscious voters are drawn from 
 an exponential distribution with mean $1$.
}\label{fig5}
\end{figure}

Finally, one may also wish to test the method against the possibility of 
 ignorant voters. To model this, we assume there is only a certain proportion
 $C$ of the raters voting according to distribution (\ref{V.Laplace}),
 while the rest 
 of the raters are voting randomly between $\epsilon$ and $10+\epsilon$.
 Figure \ref{fig5} documents the test using a population of $400$ 
 voters in a self-evaluating community. Our method provides appreciable
 improvement over the simple average 

\section{Outlook and Concluding Summary}
We have proposed an iterative refinement method to estimate the hidden intrinsic
 qualities of a set
of objects that have been evaluated by a group of raters. The method 
consists of aggregating these evaluations in a weighted sum, with the aim 
to give more weight to expert raters. Weights and qualities are
estimated iteratively from the same data set.
Extensive simulation results show that the proposed method is able to
recover the hidden attributes with remarkably good precision,
even when the conditions of the Law of Large Numbers are not fulfilled.
In particular, it overwhelms the performance of the naive simple
average in most circumstances.

The proposed method is intended to be mainly applied to virtual
communities of all kinds, where people are allowed to express their
opinions on a particular subject --which can be a material object,
a person or even another opinion. In addition to the objects' 
 intrinsic values, 
the method allows detection of the rating capabilities of the users. This
provides valuable information, since it defines
reputations~\cite{manifesto} without the need of any other feedback.
 It may constitute a
strong incentive for users to rate accurately the desired object and a
strong deterrent against cheating.

In fact the proposed method is robust against gaming. 
It remains effective in decoding the hidden
attributes in the face of two types of noise: random and intentional. 
Although we fully anticipate this approach to be effective
  when working on real data where the intrinsic values $\{Q_l \}$ 
 are not known at all, we are in the process of making  
 more critical assessments by gathering data
 from existing web sites and by even designing special purpose 
 web sites to acquire custom data. With more information shared
 and less cheating allowed, we hope our method, once implemented, can
 help our society to grow into a happier and healthier whole.

\begin{ack}
\small YKY wishes to thank Drs. John Wilbur and John Spouge for comments and
 Dr. Timothy Doerr for a critical reading of the manuscript.   
 This work was partially supported by the Intramural Research Program of
 the National Library of Medicine at National Institutes of Health/DHHS
  and by the Swiss National Science Foundation
  through project number 2051-67733. 
\end{ack}

\section*{Appendix}
In this appendix, we will derive the formula shown in eq. (\ref{I_bg}).
Consider we have $M$ objects, labeled by $O_1,~O_2,~\ldots,O_M$, 
 and we randomly put them into an ordered array of size $M$. 
 Apparently, there are $M!$ ways to do so. The question is then to obtain
 the probabilities $P(k|\ell)$ 
 for $k$ objects out of $O_1,~O_2,~\ldots,O_\ell$  to be in 
 the top $\ell$ entries of the array. 
 
When $k=\ell$ we only have $\ell !$ ways to order these $\ell$ 
 objects and $(M-\ell)!$ ways to arrange the rest. Therefore, when $k = \ell$, 
 we have $P(\ell|\ell) = (M-\ell)! \ell ! / M! $. 
 When $k < \ell$, we have to put $(\ell -k)$ objects in the lower $M-\ell$ bins
 and $k$ objects in the top $\ell$ bins. 
 There are $(M-\ell)!/(M-2\ell+k)!$ ways for the former and 
$\ell! /(\ell -k)!$ ways for the latter. Further, 
 there are 
$
{ \ell !\over k! (\ell-k)!} \equiv C_k^\ell = C_{\ell-k}^\ell
$
 ways  to choose which $k$ objects to put in the top $\ell$ bins. 
 Consequently, we have 
\bea
P(k|\ell) &=& {(M-\ell)! \over M!} 
 {\ell ! \over (\ell -k)!} C_k^\ell {(M-\ell)! \over (M-2\ell+k)!} \nonumber \\
 &=& {1\over C_\ell^M} C_k^\ell C_{\ell-k}^{M-\ell}.
\eea
As a simple check, we can verify that $\sum_{k=0}^\ell P(k|\ell) =1 $
 because 
$
\sum_{k=0}^\ell C_k^\ell C_{\ell-k}^{M-\ell} = C_\ell^M 
$
which can be easily proved by evaluating the coefficient of 
$x^\ell$ in the two equivalent expressions
 $(1+x)^M$ and $(1+x)^{M-\ell} \times (1+x)^\ell$.

It is instructive to compute the expectation value of 
$k/\ell$ for  a given $\ell$
\beas
\Avg{k\over \ell} &=& 
\sum_{k=0}^\ell {k\over \ell} C_k^\ell C_{\ell-k}^{M-\ell} 
 = \sum_{k=1}^{\ell} C_{k-1}^{\ell-1} 
 C_{(\ell-1)-(k-1)}^{(M-1)-(\ell-1)} \nonumber \\
 &=& \sum_{k'=0}^{\ell-1} C_{k'}^{\ell-1} C_{k'-(\ell-1)}^{(M-1)-(\ell-1)}
  = C_{\ell-1}^{M-1} 
\eeas
Now the quantity of interest $I(p)$, averaged under a random ensemble,
 can be expressed as 
\be
\langle I(p) \rangle_0 = {1\over [pM]} \sum_{\ell=1}^{[pM]} 
{C_{\ell-1}^{M-1}  \over C_\ell^M } = {1\over [pM]}\sum_{\ell=1}^{[pM]}
 {\ell \over M} = {[pM]+1 \over 2M}
\ee




\end{document}